\documentclass[12pt,a4paper]{article}
\usepackage{amsfonts,latexsym}
\usepackage{amsmath,amssymb}
\usepackage{graphicx,color}

\renewcommand{\thefootnote}{\fnsymbol{footnote}}

\begin{document}

\vspace{12mm}

\begin{center}
{{{\Large {\bf Black holes in massive conformal gravity }}}}\\[10mm]

{Yun Soo Myung\footnote{e-mail address: ysmyung@inje.ac.kr}}\\[8mm]

{Institute of Basic Sciences and Department  of Computer Simulation, Inje University Gimhae 621-749, Korea\\[0pt]}

\end{center}
\vspace{2mm}

\begin{abstract}
We  analyze   the classical stability of Schwarzschild black hole in
massive conformal gravity which was recently proposed for another
massive gravity model. This model in the Jordan frame is conformally
equivalent to the Einstein-Weyl gravity in the Einstein frame. The
coupled linearized Einstein equation  is decomposed into the
traceless and trace equation when one chooses $6m^2\varphi=\delta
R$. Solving the traceless  equation exhibits unstable modes
featuring the Gregory-Laflamme $s$-mode instability of
five-dimensional black string, while we find  no unstable modes when
solving the trace equation. It is shown that the instability of the
black hole in massive conformal gravity arises from the massiveness
where the geometry of extra dimension trades for mass.
\end{abstract}
\vspace{5mm}



\vspace{1.5cm}

\hspace{11.5cm}{Typeset Using \LaTeX}
\newpage
\renewcommand{\thefootnote}{\arabic{footnote}}
\setcounter{footnote}{0}


\section{Introduction}
Recently, massive conformal gravity was proposed as another massive
gravity model~\cite{Faria:2013hxa}. This model is composed of a
conformally coupled scalar to Einstein-Hilbert term (conformal
relativity) and Weyl-squared term which are invariant under
conformal transformations. Apparently, this model is related  to the
Einstein-Weyl gravity of
$R+aC_{\mu\nu\rho\sigma}C^{\mu\nu\rho\sigma}$\cite{Lu:2011ks}, which
is not manifestly invariant under conformal transformations. This
model seems to be promising because the conformal symmetry restricts
the number of counter-terms arising from the perturbative
quantization of the metric tensor~\cite{tHooft:2011aa}. However,
Stelle has shown that a definite combination of
$aC_{\mu\nu\rho\sigma}C^{\mu\nu\rho\sigma}+bR^2$ is necessary  to
improve the perturbative properties of Einstein
gravity~\cite{Stelle:1976gc}. In this sense, massive conformal
gravity including the Weyl-squared term only might not be a
candidate for a proper quantum gravity model.

On the other hand, massive conformal gravity plays a role of being
massive gravity model~\cite{deRham:2010ik,deRham:2010kj} because it
includes Einstein-Hilbert term and Weyl-squared term in addition to
a conformally coupled scalar. Actually, this action in the Jordan
frame is conformally equivalent to the Einstein-Weyl action in the
Einstein frame.   It turned out that the Schwarzschild black hole is
unstable against the Gregory-Laframme (GL) $s(l=0)$-mode metric
perturbation~\cite{Gregory:1993vy} in massive gravity
models~\cite{Babichev:2013una,Brito:2013wya,Myung:2013doa}. This is
possible because the extra dimension in five-dimensional black
string could be replaced by the mass~\cite{Deser:2013qza}. That is,
trading geometry for mass is a plausible argument for the
instability of Schwarzschild black hole in the massive gravity. If
one takes into account the number of degrees of freedom (DOF), it is
easy to show why  the Schwarzschild  black hole is physically stable
in the Einstein gravity, whereas the Schwarzschild black hole is
unstable in massive conformal  gravity. The number of DOF of the
metric perturbation is 2 DOF in the Einstein gravity, while the
number of DOF is $6=5+1$ in massive conformal  gravity. The $s$-mode
analysis is suitable for a massive graviton with 5 DOF,  whereas 1
DOF is described by  a conformally coupled scalar (linearized Ricci
scalar) which satisfies a massive scalar equation.

In this work, we investigate  the classical stability of
Schwarzschild black hole in massive conformal gravity. The coupled
linearized Einstein equation  is decomposed into the traceless and
trace equation when one chooses $6m^2\varphi=\delta R$. Solving the
traceless  equation exhibits unstable modes featuring the GL
$s$-mode instability of five-dimensional black string, while we find
no unstable modes from solving the trace equation.  This implies
that massive conformal gravity could not provide the Schwarzschild
black hole solution.

\section{Massive conformal gravity}
We consider the action for massive conformal gravity which is
composed of conformal relativity and Weyl-squared term
~\cite{Faria:2013hxa}
\begin{eqnarray}S_{\rm MCG}=\frac{1}{32 \pi}\int d^4 x\sqrt{-g}
\Big[\alpha \Big(\phi^2R+
6\partial_\mu\phi\partial^\mu\phi\Big)-\frac{1}{m^2}C^{\mu\nu\rho\sigma}C_{\mu\nu\rho\sigma}\Big],
\label{MCGact}
\end{eqnarray}
where the Weyl tensor squared is given by
\begin{equation}
C^{\mu\nu\rho\sigma}C_{\mu\nu\rho\sigma}=2\Big(R^{\mu\nu}R_{\mu\nu}-\frac{1}{3}R^2\Big)+
(R^{\mu\nu\rho\sigma}R_{\mu\nu\rho\sigma}-4R^{\mu\nu}R_{\mu\nu}+R^2).
\end{equation}
Here the last of Gauss-Bonnet term could be neglected  because it
does not contribute to equation of motion. Also, we use the Planck
units of $c=\hbar=G=1$ and $m$ is the mass of massive spin-2
graviton. The action (\ref{MCGact}) is invariant under the conformal
transformations of
\begin{equation}
g_{\mu\nu} \to \Omega^2(x)g_{\mu\nu},~~\phi \to\Omega^{-1}\phi,
\end{equation}
where $\Omega(x)$ is an arbitrary function of the spacetime
coordinates.

From (\ref{MCGact}),  the Einstein equation is derived to be
\begin{equation} \label{equa1}
\alpha
m^2\Big[\phi^2G_{\mu\nu}+g_{\mu\nu}\nabla^2(\phi^2)-\nabla_\mu\nabla_\nu(\phi^2)+6\partial_\mu\phi\partial_\nu\phi-3(\partial\phi)^2g_{\mu\nu}\Big]-2W_{\mu\nu}=0,
\end{equation}
where the Einstein tensor  is given by \begin{equation}
G_{\mu\nu}=R_{\mu\nu}-\frac{1}{2} Rg_{\mu\nu}
\end{equation}
and the Bach tensor $W_{\mu\nu}$  takes the form
\begin{eqnarray} \label{equa2}
W_{\mu\nu}&=& 2 \Big(R_{\mu\rho\nu\sigma}R^{\rho\sigma}-\frac{1}{4}
R^{\rho\sigma}R_{\rho\sigma}g_{\mu\nu}\Big)-\frac{2}{3}
R\Big(R_{\mu\nu}-\frac{1}{4} Rg_{\mu\nu}\Big) \nonumber \\
&+&
\nabla^2R_{\mu\nu}-\frac{1}{6}\nabla^2Rg_{\mu\nu}-\frac{1}{3}\nabla_\mu\nabla_\nu
R.
\end{eqnarray}
Its trace is zero  ($W^\mu~_\mu=0$).

The other scalar equation is given by
\begin{equation} \label{scalar-eq}
\nabla^2\phi-\frac{1}{6}R\phi=0,
\end{equation}
which is conformally covariant.  Taking the trace of (\ref{equa1})
leads to
\begin{equation}
-\phi^2R+3\nabla^2(\phi^2)-6(\partial \phi)^2=0 \end{equation} which
vanishes when one uses the scalar equation (\ref{scalar-eq}).

  Considering the background ansatz
\begin{equation}
\bar{R}_{\mu\nu}=0,~~\bar{R}=0,~~\bar{\phi}=\sqrt{\frac{2}{\alpha}},
\end{equation}
Eq. (\ref{equa1}) and (\ref{scalar-eq}) provide   the Schwarzschild
black hole solution \begin{equation} \label{schw} ds^2_{\rm
S}=\bar{g}_{\mu\nu}dx^\mu
dx^\nu=-f(r)dt^2+\frac{dr^2}{f(r)}+r^2d\Omega^2_2
\end{equation}
with the metric function \begin{equation} \label{num}
f(r)=1-\frac{r_0}{r}.
\end{equation}
It is easy to show that the Schwarzschild  black hole (\ref{schw})
is also the solution to the Einstein equation of $G_{\mu\nu}=0$ in
Einstein gravity.

We  introduce the metric and scalar perturbations around the
Schwarzschild  black hole
\begin{eqnarray} \label{m-p}
g_{\mu\nu}=\bar{g}_{\mu\nu}+h_{\mu\nu},~~\phi=\bar{\phi}(1+\varphi)=\sqrt{\frac{2}{\alpha}}(1+\varphi).
\end{eqnarray}
Then, the linearized Einstein equation takes the form
\begin{eqnarray} \label{lin-eq}
&&m^2\Big[\delta G
_{\mu\nu}+2\Big(\bar{g}_{\mu\nu}\bar{\nabla}^2-\bar{\nabla}_\mu\bar{\nabla}_\nu\Big)\varphi\Big]
\\ \nonumber
&& =\Big[\bar{\nabla}^2\delta
G_{\mu\nu}+2\bar{R}_{\rho\mu\sigma\nu}\delta G^{\rho\sigma}\Big]
+\frac{1}{3}\Big[\bar{g}_{\mu\nu}\bar{\nabla}^2-\bar{\nabla}_\mu\bar{\nabla}_\nu
\Big] \delta R,
\end{eqnarray}
where the linearized Einstein tensor, Ricci tensor, and Ricci scalar
are given by
\begin{eqnarray}
\delta G_{\mu\nu}&=&\delta R_{\mu\nu}-\frac{1}{2} \delta
R\bar{g}_{\mu\nu},
\label{ein-t} \\
\delta
R_{\mu\nu}&=&\frac{1}{2}\Big(\bar{\nabla}^{\rho}\bar{\nabla}_{\mu}h_{\nu\rho}+
\bar{\nabla}^{\rho}\bar{\nabla}_{\nu}h_{\mu\rho}-\bar{\nabla}^2h_{\mu\nu}-\bar{\nabla}_{\mu}
\bar{\nabla}_{\nu}h\Big), \label{ricc-t} \\
\delta R&=& \bar{g}^{\mu\nu}\delta R_{\mu\nu}= \bar{\nabla}^\mu
\bar{\nabla}^\nu h_{\mu\nu}-\bar{\nabla}^2 h \label{Ricc-s}
\end{eqnarray}
with $h=h^\rho~_\rho$.

From (\ref{scalar-eq}),  we derive the linearized scalar equation
\begin{equation} \label{linsca}
\bar{\nabla}^2\varphi-\frac{1}{6}\delta R=0
\end{equation}
which is surely a coupled equation for $\varphi$ and $\delta R$.
Plugging (\ref{linsca}) into (\ref{lin-eq}), one finds a simpler
linearized Einstein equation
\begin{eqnarray} \label{silin-eq}
&&m^2\Big[\delta G
_{\mu\nu}-2\bar{\nabla}_\mu\bar{\nabla}_\nu\varphi\Big]  \nonumber
\\
&&=\Big[\bar{\nabla}^2\delta
G_{\mu\nu}+2\bar{R}_{\rho\mu\sigma\nu}\delta
G^{\rho\sigma}\Big]+\frac{1}{3}\Big[\bar{g}_{\mu\nu}(\bar{\nabla}^2-m^2)-\bar{\nabla}_\mu\bar{\nabla}_\nu
\Big] \delta R.
\end{eqnarray}

It might be difficult to solve (\ref{silin-eq}) directly because it
is a coupled second-order equation for $\delta G_{\mu\nu}$, $\delta
R$, and $\varphi$. Taking the trace of (\ref{silin-eq}) together
with $\delta G^\mu~_\mu=-\delta R$ leads to (\ref{linsca}) too.
 In order to simplify the linearized equation
(\ref{silin-eq}), one way is to find  a condition of non-propagating
linearized Ricci scalar ($\delta R=0$). However, it is not justified
to impose $\delta R=0$ because of conformal symmetry in massive
conformal gravity. In Appendix, we have $\delta R=0$ for the new
massive conformal gravity where the conformal symmetry is broken due
to the addition  of the Einstein-Hilbert term.

The other way to resolve the coupling difficulty  is to propose a
relation between $\varphi$ and $\delta R$ because the massive
conformal gravity implies 6 DOF of massive graviton (with 5 DOF) and
scalar. If one requires the relation \begin{equation} \label{rrs}
\varphi=\frac{1}{6m^2} \delta R
\end{equation}
the linearized equation (\ref{silin-eq})[(\ref{lin-eq})] is
simplified further as
\begin{eqnarray} \label{simlin-eq}
\bar{\nabla}^2\delta G_{\mu\nu}+2\bar{R}_{\rho\mu\sigma\nu}\delta
G^{\rho\sigma}=m^2\delta G_{\mu\nu}
 \end{eqnarray}
Before we proceed, we would like to mention that the relation
(\ref{rrs}) between $\varphi$ and $\delta R$ is taken specially for
the massive conformal gravity. If we do not use this relation, we
could not make a further progress on the stability analysis. The
apparently two DOF of $\varphi$ and $\delta R$ becomes a single DOF
due to the relation (\ref{rrs}).
 Plugging
(\ref{rrs}) into the linearized scalar equation (\ref{linsca}) leads
to the massive scalar equation
\begin{equation} \label{masca}
\Big(\bar{\nabla}^2-m^2\Big)\varphi=0.
\end{equation}
Also, the same equation is recovered when one takes the trace of
(\ref{simlin-eq})
\begin{equation} \label{maricc}
\Big(\bar{\nabla}^2-m^2\Big)\delta R=0
\end{equation}
which is called the trace equation.  However, Eq. (\ref{simlin-eq})
describes 6 DOF of a massive graviton and (Ricci) scalar wholly.
Splitting $\delta R_{\mu\nu}$ into the traceless linearized Ricci
tensor $\delta \tilde{R}_{\mu\nu}$ with $\bar{g}^{\mu\nu}\delta
\tilde{R}_{\mu\nu}=0$ and the linearized Ricci scalar $\delta R$ as
\begin{equation}
\delta R_{\mu\nu}=\delta \tilde{R}_{\mu\nu}+\frac{1}{4}\delta R
\bar{g}_{\mu\nu},
\end{equation}
the linearized Einstein tensor is given by
\begin{equation}
\delta G_{\mu\nu}=\delta \tilde{R}_{\mu\nu}-\frac{1}{4}\delta R
\bar{g}_{\mu\nu}.
\end{equation}
Then, the linearized Einstein equation (\ref{simlin-eq}) takes the
form
\begin{eqnarray} \label{fein-eq}
\bar{\nabla}^2\delta
\tilde{R}_{\mu\nu}+2\bar{R}_{\rho\mu\sigma\nu}\delta
\tilde{R}^{\rho\sigma}-m^2\delta
\tilde{R}_{\mu\nu}-\frac{\bar{g}_{\mu\nu}}{4}\Big(\bar{\nabla}^2-m^2\Big)\delta
R=0.
\end{eqnarray}
At first sight, Eq. (\ref{fein-eq}) seems to be  a coupled equation
for $\delta \tilde{R}_{\mu\nu}$ and $\delta R$. Using the trace
equation (\ref{maricc}) to eliminate $\delta R$, Eq. (\ref{fein-eq})
is reduced to the traceless linearized Ricci tensor equation
\begin{eqnarray} \label{tlricci-eq} \bar{\nabla}^2\delta
\tilde{R}_{\mu\nu}+2\bar{R}_{\rho\mu\sigma\nu}\delta
\tilde{R}^{\rho\sigma}=m^2\delta \tilde{R}_{\mu\nu}
\end{eqnarray}
which is our main result.

 On the other hand, we note that in the
Einstein-Weyl gravity~\cite{Myung:2013doa}, the non-propagation of
the linearized Ricci scalar ($\delta R=0$) is an essential
requirement to arrive at the linearized massive Ricci tensor
equation
\begin{eqnarray} \label{ricci-eq}
\bar{\nabla}^2\delta R_{\mu\nu}+2\bar{R}_{\rho\mu\sigma\nu}\delta
R^{\rho\sigma}-m^2\delta R_{\mu\nu}=0
\end{eqnarray}
which describes a massive graviton with 5 DOF propagating around the
Schwarzschild black hole.

However, the massive conformal gravity implies that  the linearized
Einstein tensor with 6 DOF ($\delta \tilde{R}_{\mu\nu}$ and $\delta
R$) propagating the Schwarzschild black hole satisfies the traceless
equation (\ref{tlricci-eq}) and  the trace equation (\ref{maricc}).
The traceless condition of $\delta G^\mu~_\mu=-\delta R=0$ is an
important requirement to show the GL instability of the
Schwarzschild black hole and it could be  achieved only in the
Einstein-Weyl gravity.  On the contrary, the trace equation
(\ref{maricc}) plays an important role of obtaining the traceless
equation (\ref{tlricci-eq}) in massive conformal gravity.  In the
next section, we will prove that the conformally invariant action
(\ref{MCGact})  in the Jordan frame is conformally equivalent to the
Einstein-Weyl action in the Einstein frame.

\section{Massive conformal gravity in Einstein frame}
In this section, we transform the conformally invariant action
(\ref{MCGact}) into the corresponding action in the Einstein frame.
First of all, it would be better to show that the conformally
invariant action (\ref{MCGact}) is nothing but the $\omega=-3/2$
Brans-Dicke theory plus Weyl-squared term for $\alpha=1/6$ when one
chooses ~\cite{Dabrowski:2005yn}
\begin{equation}
\frac{1}{12}\phi^2=e^{-\Phi}.
\end{equation}
Then,  (\ref{MCGact}) is given by \begin{equation} \label{tMCG}
\tilde{S}_{\rm MCG}=\frac{1}{16\pi}\int
d^4x\sqrt{-g}\Big[e^{-\Phi}\Big(R+\frac{3}{2}\partial_\mu\Phi\partial^\mu\Phi\Big)-\frac{1}{2m^2}C^{\mu\nu\rho\sigma}C_{\mu\nu\rho\sigma}\Big].
\end{equation}
On the other hand, the Brans-Dicke theory plus Weyl-squared term is
described by
\begin{equation} \label{BDA}
S_{BDW}^\omega=\frac{1}{16\pi}\int d^4x \sqrt{-g}\Big[\phi_{\rm
BD}R-\frac{\omega}{\phi_{\rm BD}}\partial_\mu \phi_{\rm
BD}\partial^\mu \phi_{\rm
BD}-\frac{1}{2m^2}C^{\mu\nu\rho\sigma}C_{\mu\nu\rho\sigma}\Big].
\end{equation}
Choosing $\phi_{\rm BD}=e^{-\Phi}$, (\ref{BDA}) could be rewritten
as
\begin{equation} \label{tBDA}
\tilde{S}_{\rm BDW}^\omega=\frac{1}{16\pi}\int
d^4x\sqrt{-g}\Big[e^{-\Phi}\Big(R-\omega\partial_\mu\Phi\partial^\mu\Phi\Big)-\frac{1}{2m^2}C^{\mu\nu\rho\sigma}C_{\mu\nu\rho\sigma}\Big].
\end{equation}
We note  that $\tilde{S}_{\rm BDW}^{\omega=-3/2}=\tilde{S}_{\rm
MCG}$ (\ref{tMCG}), indicating that  the conformal relativity is
just  the Brans-Dicke theory with $\omega=-3/2$ in the Jordan frame.

Now we make conformal transformation of the conformally invariant
action (\ref{MCGact}) with $\alpha=1/6$ only  by
choosing~\cite{Farajollahi:2010ni,Romero:2012hs}
\begin{equation}
\hat{g}_{\mu\nu}=\Omega^2
g_{\mu\nu},~~\hat{\phi}=\phi-\phi=0,~~\Omega=\frac{\phi}{2\sqrt{3}}.
\end{equation}
Then, the transformed action takes the form
\begin{eqnarray}\hat{S}_{\rm MCG}=\frac{1}{16 \pi}\int d^4 x\sqrt{-\hat{g}}
\Big[\hat{R}-\frac{1}{2m^2}\hat{C}^{\mu\nu\rho\sigma}\hat{C}_{\mu\nu\rho\sigma}\Big]
\label{MCGactt}
\end{eqnarray}
which is noting but the Einstein-Weyl gravity in the Einstein frame.
Hence it is clear  that   the conformally invariant action
(\ref{MCGact}) ($\omega=-3/2$ Brans-Dicke theory plus Weyl-squared
term) in the Jordan frame is conformally equivalent to the
Einstein-Weyl action (\ref{MCGactt}) in the Einstein frame.  The
Schwarzschild black hole (\ref{schw}) is also obtained as the
solution to the Einstein equation. Its linearized Einstein equation
is given by~\cite{Myung:2013doa}
\begin{eqnarray} \label{einfricci-eq} \bar{\nabla}^2\delta
\hat{R}_{\mu\nu}+2\bar{R}_{\rho\mu\sigma\nu}\delta
\hat{R}^{\rho\sigma}-m^2\delta \hat{R}_{\mu\nu}=0
\end{eqnarray}
together with transverse-traceless condition of
$\bar{\nabla}^\mu\delta \hat{R}_{\mu\nu}=0$ and $\delta \hat{R}=0$.
This implies that even though a conformally coupled scalar $\phi$
provides a different linearized Einstein equation (\ref{simlin-eq})
with (\ref{maricc}) in the Jordan frame, it disappears  in the
Einstein frame.

If one starts with a non-conformally invariant action, there exists
a scalar kinetic term of $-\frac{\lambda}{2}
\hat{g}^{\mu\nu}\partial_\mu\phi\partial_\nu\phi$ which could be
reduced to a canonical form of $-\frac{1}{2}
\hat{g}^{\mu\nu}\partial_\mu\Psi\partial_\nu\Psi$ in terms of a
minimally coupled scalar $\Psi=\sqrt{\lambda}\phi$.  The
non-conformally invariant action ($\omega>-3/2$ Brans-Dicke theory
plus Weyl-squared term) in the Jordan frame is conformally
equivalent to the scalar-Einstein-Weyl gravity in the Einstein
frame~\cite{Faraoni:1999hp}
\begin{eqnarray}\hat{S}_{\rm MNCG}=\frac{1}{16 \pi}\int d^4 x\sqrt{-\hat{g}}
\Big[\hat{R}-\frac{1}{2}
\hat{g}^{\mu\nu}\partial_\mu\Psi\partial_\nu\Psi-\frac{1}{2m^2}\hat{C}^{\mu\nu\rho\sigma}\hat{C}_{\mu\nu\rho\sigma}\Big].
\label{Einact}
\end{eqnarray}
It was proposed that the stability of black holes does not depend on
the frame~\cite{Tamaki:2003ah}, even though there exists an apparent
difference between (\ref{simlin-eq}) and (\ref{einfricci-eq}). The
difference is the trace equation (\ref{maricc}) which becomes
 the conformal scalar equation (\ref{masca}). We will check the
 above proposal.

\section{Instability of Schwarzschild  black hole
in massive conformal  gravity}
 Considering  the number of
DOF, it is helpful to show why  the Schwarzschild black hole is
physically stable in the Einstein
gravity~\cite{Regge:1957td,Zeri,Vish}, whereas the Schwarzschild
black hole is unstable in massive conformal  gravity. From Eq.
(\ref{simlin-eq}) together with the linearized Bianchi identity
($\bar{\nabla}^\mu\delta G_{\mu\nu}=0$), the number of DOF for
massive spin-2 graviton is $10-4=6$ in massive conformal gravity. On
the other hand, the number of DOF for the massless spin-2 graviton
is 2 in  Einstein gravity since one requires $-4$ further for a
residual diffeomorphism after a gauge-fixing and the traceless
condition.   The $s$-mode analysis is relevant to the massive
graviton in massive conformal gravity, but not to the massless
graviton in the Einstein gravity. In general, the $s$-mode analysis
of the massive graviton with $5$ DOF shows the GL-instability which
never appears in the massless spin-2 analysis.

To perform the stability of Schwarzschild black hole in massive
conformal gravity completely, we have to solve two linearized
equations: the trace equation (\ref{maricc}) and traceless equation
(\ref{tlricci-eq}) with the same mass-squared $m^2$. These are
different from those arising from the forth-order gravity of
$R-\alpha R^2-\beta R_{\mu\nu}R^{\mu\nu}$~\cite{Myung:2013doa}
because the latter provides different masses
$m_0^2=-1/2(3\alpha+\beta)$ and $m^2_2=1/\beta$. If
$\alpha=-\beta/3$(Weyl-squared term), the linearized  Ricci scalar
is decoupled from the theory because its mass $m^2_0$ blows up.

First of all, we wish to solve  the massive scalar equation
(\ref{masca}) [equivalently, Ricci scalar equation (\ref{maricc})]
around the
 Schwarzschild black hole. It turned out that the scalar mode does
 not have any unstable modes if $m^2\ge 0$~\cite{Kwon:1986dw,Myung:2011ih}. Explicitly, considering the  scalar
 perturbation
 \begin{equation}
\varphi(t,r,\theta,\phi)=e^{i\omega t}\frac{\psi(r)}{r}
 Y_{lm}(\theta,\phi)
\end{equation}
and introducing the tortoise coordinate \begin{equation}
r^*=r+r_0\ln\Big[\frac{r}{r_0}-1\Big] \end{equation}
 the linearized
equation (\ref{masca}) reduces to the Schr\"odinger-type equation as
\begin{equation}
\frac{d^2\psi}{dr^{*2}}+(\omega^2-V_{\psi})\psi=0
\end{equation}
with the potential
\begin{equation}
V_{\psi}=\Big(1-\frac{r_0}{r}\Big)\Big[\frac{l(l+1)}{r^2}+\frac{r_0}{r^3}+m^2\Big].
\end{equation}
The potential $V_{\psi}$  is always positive exterior the event
horizon $r=r_0$ for $l\ge 0$ and $m^2\ge0$, implying that the black
hole is stable against the  scalar [Ricci scalar] perturbation.

 However, the
$s$-mode  analysis is responsible  for detecting  an instability of
a massive graviton propagating on the Schwarzschild  black hole in
massive  gravity. The even-parity metric perturbation is designed
for a $s(l=0)$-mode analysis in the massive gravity and whose form
is given by $H_{tt},~H_{tr},~H_{rr},$ and $K$
as~\cite{Gregory:1993vy}
\begin{eqnarray}
h^{(m)}_{\mu\nu}=e^{\Omega t} \left(
\begin{array}{cccc}
H_{tt}(r) & H_{tr}(r) & 0 & 0 \cr H_{tr}(r) & H_{rr}(r) & 0 & 0 \cr
0 & 0 &  K(r) & 0 \cr 0 & 0 & 0 & \sin^2\theta K(r)
\end{array}
\right). \label{evenp}
\end{eqnarray}
Even though one starts with 4 DOF, they are  related to each other
when one uses the transverse-traceless gauge of $\bar{\nabla}^\mu
h^{(m)}_{\mu\nu}=0$ and $h^{(m)}=0$. Hence, we  have one decoupled
equation for  $H_{tr}$  from the massive graviton equation
\begin{eqnarray} \label{h-eq}
\bar{\nabla}^2h^{(m)}_{\mu\nu}+2\bar{R}_{\rho\mu\sigma\nu}h^{(m)\rho\sigma}=m^2h^{(m)}_{\mu\nu}.
\end{eqnarray}
Since Eq.(\ref{h-eq}) is the same linearized equation for
four-dimensional metric perturbation around five-dimensional black
string, we use the GL instability analysis  in asymptotically flat
spacetimes~\cite{Gregory:1993vy}. Eliminating all but $H_{tr}$,
Eq.(\ref{h-eq}) reduces to a second-order radial equation for
$H_{tr}$
\begin{equation} \label{second-eq} A
H_{tr}^{''} +B H_{tr}^{'}+CH_{tr}=0,
\end{equation}
where $A,B$ and $C$ are given by
\begin{eqnarray}
&&A=-m^2
f-\Omega^2+\frac{f^{'2}}{4}-\frac{ff^{''}}{2}-\frac{ff^{'}}{r},
\end{eqnarray}
\begin{eqnarray}
&&
B=-2m^2f^{'}-\frac{3f^{'}f^{''}}{2}-\frac{3\Omega^2f^{'}}{f}+\frac{3f^{'3}}{4f}+\frac{2m^2f}{r}+\frac{2\Omega^2}{r}+\frac{3f^{'2}}{2r}
+\frac{ff^{''}}{r} -\frac{2ff^{'}}{r^2},
\end{eqnarray}
\begin{eqnarray}
C&=&m^4+\frac{\Omega^4}{f^2}+\frac{2m^2\Omega^2}{f}-\frac{5\Omega^2f^{'2}}{4f^2}+\frac{m^2f^{'2}}{4f}+\frac{f^{'4}}{4f^2}-\frac{m^2f^{''}}{2}-\frac{\Omega^2f^{''}}{2f}-\frac{f^{'2}f^{''}}{4f}-\frac{f^{''2}}{2}
\nonumber\\
&&-\frac{2m^2f^{'}}{r}-\frac{\Omega^2f^{'}}{r f}+\frac{f^{'3}}{r
f}-\frac{3f^{'}f^{''}}{r}
+\frac{2\Omega^2}{r^2}+\frac{2m^2f}{r^2}-\frac{5f^{'2}}{2r^2}+\frac{ff^{''}}{r^2}+\frac{2ff^{'}}{r^3}
\end{eqnarray}
with the metric function $f=1-r_0/r$ (\ref{num}).

 It is worth noting that the
$s$-mode perturbation is described by single DOF but not 5 DOF.  We
solve   (\ref{second-eq}) numerically and find  unstable modes. See
Fig. 1 that is  generated from the numerical analysis.  From the
observation of Fig. 1 with ${\cal O}(1)\simeq 0.86$, we find
unstable modes~\cite{Babichev:2013una} for
\begin{equation} \label{unst-con}
0<m<\frac{{\cal O}(1)}{r_0} \end{equation} with  mass $m$.
\begin{figure*}[t!]
   \centering
   \includegraphics{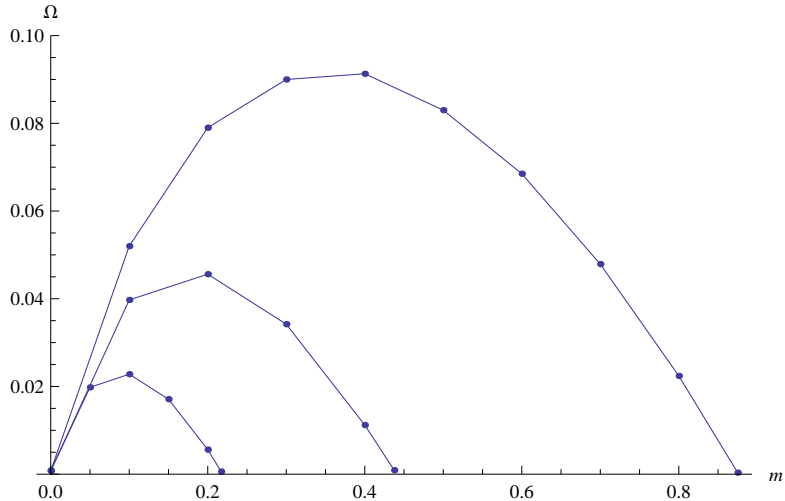}
\caption{Plots of unstable modes on three curves with $r_0=1,2,4$.
The $y(x)$-axis denote $\Omega(m)$. The smallest curve represents
$r_0=4$, the medium denotes $r_0=2$, and the largest one shows
$r_0=1$.}
\end{figure*}
As the horizon size $r_0$ increases, the instability becomes weak as
in the Schwarzschild black hole.

For a massive gravity theory in the Minkowski background, there is
correspondence between linearized Ricci tensor $\delta R_{\mu\nu}$
and Ricci spinor $\Phi_{ABCD}$ when one uses  the Newman-Penrose
formalism~\cite{Newman:1961qr}.  Here the
 massive gravity  requires null complex tetrad to
specify six polarization modes~\cite{Eardley:1974nw,Moon:2011gg}.
This implies that in massive conformal gravity, one takes  the
linearized Ricci tensor $\delta R_{\mu\nu}$ (\ref{ricc-t}) with 5
DOF as physical observables~\cite{Myung:2013doa} by requiring the
transversality condition of $\bar{\nabla}^\mu \delta R_{\mu\nu}=0$
from the contracted  Bianchi identity and the traceless condition of
$\delta R=0$. That is, the traceless linearized Ricci tensor $\delta
\tilde{R}_{\mu\nu}$ has the same 5 DOF as the metric perturbation
$h_{\mu\nu}$ does have in massive gravity theory.  Actually, Eq.
(\ref{tlricci-eq}) is considered as a boosted-up version of
(\ref{h-eq})~\cite{Bergshoeff:2013vra}.
 Similarly, we find
Eq.(\ref{h-eq}) when we  replace    $ \delta \tilde{R}_{\mu\nu}$ by
$h^{(m)}_{\mu\nu}$  in (\ref{tlricci-eq}). Hence, a relevant
equation for $\delta\tilde{ R}_{tr}$ takes the same form
\begin{equation} \label{secondG-eq} A
\delta \tilde{R}_{tr}^{''} +B\delta \tilde{R}_{tr}^{'}+C\delta
\tilde{R}_{tr}=0
\end{equation}
which shows the same unstable modes appeared in Fig. 1.

Consequently, we have found unstable $s$-mode from the traceless
equation (\ref{tlricci-eq}), but have not found unstable modes from
the trace equation (\ref{maricc}) [scalar equation (\ref{masca})] in
the Jordan frame. If one uses the linearized equation
(\ref{einfricci-eq}) arisen from the Einstein-weyl gravity in the
Einstein frame, one finds the same unstable modes. This implies
that the instability of black holes in massive gravity does not
depend on the frame.

\section{Discussions}
We discuss on the following issues.

\noindent $\bullet$ Ghosts and linearized Ricci tensor \\
Since the linearized equation (\ref{lin-eq}) is a fourth-order
derivative equation, it involves the linearized
ghosts~\cite{Bergshoeff:2013vra}. The ghost appears surely when one
introduces an auxiliary tensor $f_{\mu\nu}$ to reduce fourth-order
gravity theory to second-order theory~\cite{Hyun:2011ej}. This
implies that if one uses the massive spin-2 equation (\ref{h-eq}) to
analyze the instability of Schwarzschild black hole in the massive
conformal gravity, its instability  might not be legitimate. If one
uses the linearized Ricci tensor $\delta \tilde{R}_{\mu\nu}$ instead
of the metric perturbation $h_{\mu\nu}$~\cite{Myung:2013doa}, its
linearized equation is a second-order equation
(\ref{tlricci-eq}) which is free from any ghosts. \\
\noindent $\bullet$ Renormalizability and conformal symmetry  \\
It was suggested  that the conformal invariant action (\ref{MCGact})
enhances the renormalizability because the conformal symmetry
restricts the number of counter-terms arising from the perturbative
quantization of the metric tensor~\cite{Faria:2013hxa}. However,
Stelle~\cite{Stelle:1976gc} has  shown that the quadratic curvature
gravity of $a(R_{\mu\nu}^2 -R^2/3)+ b R^2$ in addition to the
Einstein-Hilbert term ($R$) is necessary  to improve the
perturbative properties of Einstein gravity.   If $ab\not=0$, the
renormalizability was achieved but the unitarity was violated,
indicating that the renormalizability and unitarity exclude to each
other. Although the $a$-term of providing the massive graviton
improves the ultraviolet divergence, it induces ghost excitations
which spoil the unitarity simultaneously.  The price one has to pay
for making the theory renormalizable in this way is the loss of
unitarity. If one excludes $bR^2$, there is no massive spin-0
corrections. In this sense,  the conformal invariant action
(\ref{MCGact}) is unhealthy and it might not enhance the
renormalizability without unitarity. \\
\noindent $\bullet$ Massive conformal gravity and black hole  \\
As was shown in most  massive gravity
theories~\cite{Babichev:2013una,Brito:2013wya,Myung:2013doa}, it is
difficult for massive conformal gravity to accommodate the static
black hole solution because the GL $s(l=0)$-mode
instability~\cite{Gregory:1993vy} was found. It could be understood
that the instability of the black hole in massive conformal gravity
arises from the massiveness of $m^2\not=0$, where the geometry of
extra dimension in five-dimensional black string
is replaced by the  mass~\cite{Deser:2013qza}. \\
\noindent $\bullet$ Role of a conformally coupled scalar $\varphi$\\
Even the scalar is conformally coupled to Einstein-Hilbert action to
give a conformally invariant action, its role in testing the black
hole stability is trivial because   the conformally invariant action
(\ref{MCGact}) ($\omega=-3/2$ Brans-Dicke theory plus Weyl-squared
term) in the Jordan frame is conformally equivalent to the
Einstein-Weyl action (\ref{MCGactt}) in the Einstein frame.  The
instability of Schwarzschild black hole is determined definitely by
the massive linearized Ricci tensor equations (\ref{tlricci-eq}) and
(\ref{einfricci-eq}) which are the same equation in both theories.
The scalar field equation (\ref{masca}) [Ricci scalar equation
(\ref{maricc}) using $\varphi=\delta R/6m^2$] did not show any
unstable modes for $m^2\ge 0$. This implies that the instability of
Schwarzschild black hole is independent of choosing a frame.
 \\
\noindent $\bullet$ $f(R)$-gravity and massive conformal gravity.
\\
A simple model of $f(R)=R+\alpha R^2$ provides a ghost-free massless
graviton and massive spin-0 graviton~\cite{Myung:2011ih}, while
massive conformal gravity shows a massless graviton, scalar, and
massive spin-2 graviton with ghosts in terms of  metric tensor. A
similarity between two gravity theories is that both have a
propagating linearized Ricci scalar ($\delta R$). A difference is
that $f(R)$ gravity does
 not provide a propagating Ricci tensor $(\delta R_{\mu\nu})$, while massive conformal gravity
have a propagating Ricci tensor.

 \vspace{1cm}

{\bf Acknowledgments}
 \vspace{1cm}

The author thanks Taeyoon Moon for helpful  discussions.  This work
was supported by the National Research Foundation of Korea (NRF)
grant funded by the Korea government (MEST)
(No.2012-R1A1A2A10040499).

\section*{Appendix: New massive conformal gravity} Adding the Einstein-Hilbert term  is an easy way
to break conformal symmetry in massive conformal
gravity~\cite{Flanagan:1996gw}. Then, the new massive conformal
gravity action is proposed  by
\begin{eqnarray}S_{\rm NMCG}=\frac{1}{32 \pi }\int d^4 x\sqrt{-g}
\Big[-R+\alpha\Big(\phi^2R+
6\partial_\mu\phi\partial^\mu\phi\Big)-\frac{1}{m^2}C^{\mu\nu\rho\sigma}C_{\mu\nu\rho\sigma}\Big].
\label{NMCG}
\end{eqnarray}
The Einstein equation is changed to be
\begin{equation} \label{nequa1}
G_{\mu\nu}=\alpha
\Big[\phi^2G_{\mu\nu}+g_{\mu\nu}\nabla^2(\phi^2)-\nabla_\mu\nabla_\nu(\phi^2)+6\partial_\mu\phi\partial_\nu\phi-3(\partial\phi)^2g_{\mu\nu}\Big]-\frac{2}{m^2}W_{\mu\nu}.
\end{equation}
However, the scalar equation remains unchanged as
\begin{equation} \label{ascalar-eq}
\nabla^2\phi-\frac{1}{6}R\phi=0.
\end{equation}
Taking the trace of (\ref{nequa1}) leads to
\begin{equation} \label{ricciz}
R=0
\end{equation}
which simplifies the scalar equation (\ref{ascalar-eq})  as the
uncoupled massless scalar equation
\begin{equation}
\nabla^2\phi=0.
\end{equation}
The linearized Einstein equation around the Schwarzschild black hole
is modified into
\begin{eqnarray} \label{nlin-eq}
&&m^2\Big[\frac{1}{2}\delta G
_{\mu\nu}+2\bar{g}_{\mu\nu}\bar{\nabla}^2\varphi-2\bar{\nabla}_\mu\bar{\nabla}_\nu\varphi\Big]
\\ \nonumber
&& =\Big[\bar{\nabla}^2\delta
G_{\mu\nu}+2\bar{R}_{\rho\mu\sigma\nu}\delta G^{\rho\sigma}\Big]
-\frac{1}{3}\Big[\bar{\nabla}_\mu\bar{\nabla}_\nu-\bar{g}_{\mu\nu}\bar{\nabla}^2
\Big] \delta R.
\end{eqnarray}
The linearized scalar equation is
\begin{equation}\label{nlsca}
\bar{\nabla}^2\varphi=0.
\end{equation}
Taking the trace of the linearized Einstein equation and using
(\ref{nlsca}), one has
\begin{equation}
-\frac{m^2}{2}\delta R=0
\end{equation}
which implies the non-propagation of linearized Ricci scalar
\begin{equation}
\delta R=0
\end{equation}
unless $m^2=0$. We note that $\delta R=0$ is confirmed from
linearizing $R=0$ (\ref{ricciz}). The choice of $\delta R=0$
reflects why we consider not the massive conformal gravity
(\ref{MCGact}) but the new massive conformal gravity (\ref{NMCG}) as
a starting action. If one does not break conformal symmetry, one
could not achieve the non-propagation of the Ricci scalar. Plugging
$\delta R=0$  and (\ref{nlsca}) into Eq. (\ref{nlin-eq}) leads to
the massive equation for the linearized Ricci
tensor~\cite{Myung:2013doa}
\begin{equation} \label{slin-eq}
\bar{\nabla}^2\delta R_{\mu\nu}+ 2\bar{R}_{\rho\mu\sigma\nu}\delta
R^{\rho\sigma}=m^2\Big[\frac{1}{2}\delta
R_{\mu\nu}-2\bar{\nabla}_\mu\bar{\nabla}_\nu\varphi\Big],
\end{equation}
which is still difficult to be solved because of  coupling $\delta
R_{\mu\nu}$ and $\varphi$.

\end{document}